\documentclass[12pt]{spieman}  
\usepackage{amsmath,amsfonts,amssymb}
\usepackage{graphicx}
\usepackage{setspace}
\usepackage{tocloft}
\usepackage{placeins}

\usepackage{wrapfig}
\usepackage{subcaption}
\usepackage{graphbox}

\usepackage[export]{adjustbox}

\usepackage[margin=1.0in, bottom=1.5in]{geometry}

\title{Optimized Spatial-Spectral CT for Multi-Material Decomposition\vspace{-1em}}

\author{Matthew Tivnan}
\author{Wenying Wang}
\author{Steven Tilley}
\author{Jeffrey H.  Siewerdsen}
\author{J. Webster Stayman}
\affil{Department of Biomedical Engineering, Johns Hopkins University, 720 Rutland Ave., Baltimore MD 21205}

\cftpagenumbersoff{figure} 
\cftpagenumbersoff{table}

\begin{document} 

\maketitle

\vspace{-8mm}

\begin{abstract}
Spectral CT is an emerging modality that uses a data acquisition scheme with varied spectral responses to provide enhanced material discrimination in addition to the structural information of conventional CT. Existing clinical and preclinical designs with this capability include kV-switching, split-filtration, and dual-layer detector systems to provide two spectral channels of projection data. In this work, we examine an alternate design based on a spatial-spectral filter. This source-side filter is made up a linear array of materials that divide the incident x-ray beam into spectrally varied beamlets. This design allows for any number of spectral channels; however, each individual channel is sparse in the projection domain. Model-based iterative reconstruction methods can accommodate such sparse spatial-spectral sampling patterns and allow for the incorporation of advanced regularization. With the goal of an optimized physical design, we characterize the effects of design parameters including filter tile order and filter tile width and their impact on material decomposition performance. We present results of numerical simulations that characterize the impact of each design parameter using a realistic CT geometry and noise model to demonstrate feasibility. Results for filter tile order show little change indicating that filter order is a low-priority design consideration. We observe improved performance for narrower filter widths; however, the performance drop-off is relatively flat indicating that wider filter widths are also feasible designs.
\end{abstract}

\begin{spacing}{1}   

\vspace{-4mm}

\section{Introduction}

\vspace{-2mm}

Spectral CT is an emerging modality which incorporates varied x-ray spectral responses into measurements to enable material decomposition based on material-specific energy dependencies. Existing dual-energy CT technologies including dual-source and kv-switching systems have already opened the door to new clinical applications including water-calcium decomposition for quantitative bone imaging and simultaneous structural-functional scans with iodine-based contrast agents.\cite{mccollough2015dual} Spectral CT systems with more than two spectral channels enable multi-contrast studies and have the potential to push the limits of low concentration sensitivity. Photon-counting CT detectors are an emerging technology that can have multiple spectral channels \cite{symons2017photon}, but they are subject to a number of limitations including lower count rates and spectral distortions. Other multi-energy options include a combination of dual-energy methods (e.g. split-filters and dual-source) \cite{primak2009improved}.

\begin{figure}
    \begin{subfigure}[t]{.45\textwidth}
      \centering
      \includegraphics[scale = 0.55]{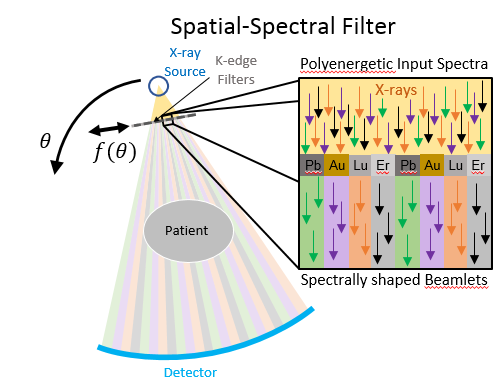}
      \caption{}
      \label{fig:system}
    \end{subfigure}%
    \begin{subfigure}[t]{.55\textwidth}
      \centering
      \includegraphics[clip, trim={5cm 1cm 6cm 1cm}, scale = 0.3]{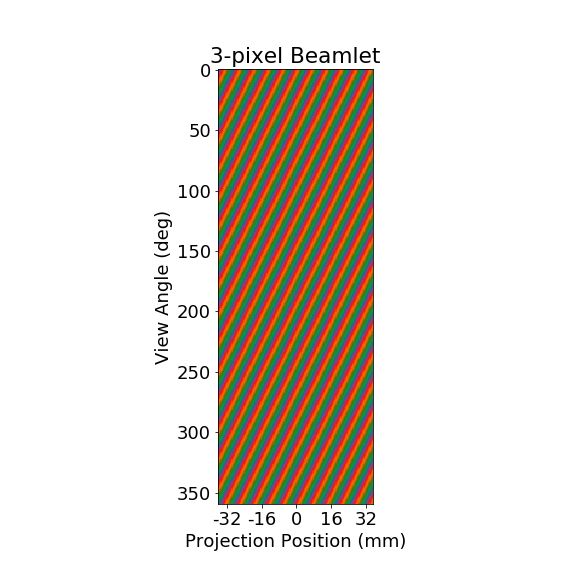}
      \includegraphics[clip, trim={5cm 1cm 6cm 1cm}, scale = 0.3]{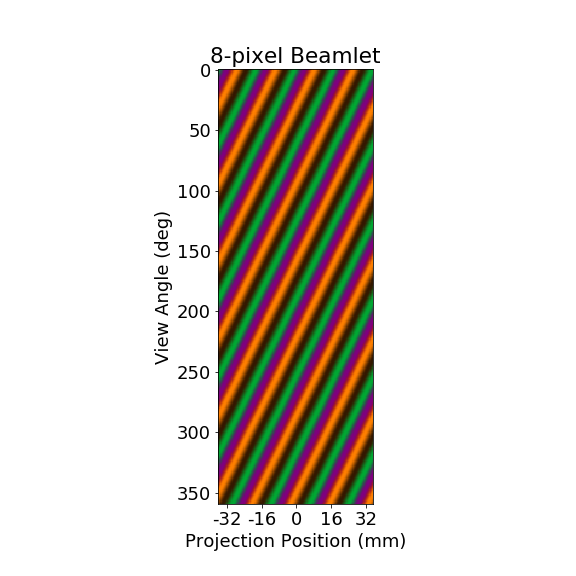}
      \includegraphics[clip, trim={5cm 1cm 6cm 1cm}, scale = 0.3]{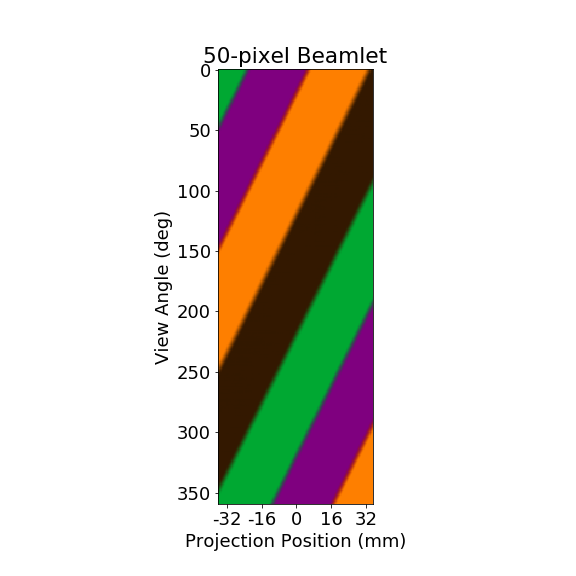}
      \caption{}
      \label{fig:spatialSpectralSampling}
    \end{subfigure}%
    \vspace{4mm}
    \caption{ (a) Schematic of a spatial-spectral CT system and (b) sampling patterns for 3 beamlet width cases. 
    The Pb, Au, Lu, and Er spectra, are green, violet, orange, and black, respectively.}
    \label{fig:spectralSampling}
\end{figure}

We have previously proposed a multi-energy CT system involving a source-side spatial-spectral filter \cite{stayman2018model} \cite{tivnan2019physical} \--- a generalization and extension of the split-filter design. The filter is composed of a tiled array of metals that divide the incident beam into spectrally varied beamlets. The filter is translated relative to the source to permit more uniform spatial-spectral sampling patterns throughout the imaging volume. (See Figure \ref{fig:system} for an illustration of a spatial-spectral CT system.) 
Each spectral channel is sparse in the projection domain, so a model-based iterative reconstruction (MBIR) is adopted for direct estimation of material density/concentration image volumes. Such sparse projection data are well-suited to compressed sensing or other advanced regularization schemes \cite{bian2010evaluation} \cite{koesters2017sparsect}.  Spatial-spectral filters permit flexibility in the number of spectra that can be incorporated, and can potentially be combined with other spectral methods for improved decomposition performance. 

In this work, we seek optimized designs for physical implementation by characterizing the impact of filter parameters, including filter tile width and filter tile order, on material decomposition performance. Simulation studies are presented and practical design constraints are identified for the new spectral CT system. By modeling a realistic CT geometry and noise properties including low concentrations of three different contrast agents in water, we intend to demonstrate the capability of this new design to explore the limits of material decomposition with spatial-spectral CT.  


\vspace{-1mm}

\section{Methods}

\vspace{-2mm}

\subsection{Numerical Phantom and Simulated Acquisition}

\vspace{-2mm}

The numerical CT phantom used for simulation experiments is shown in Figure \ref{fig:x_GT}. The object consists of four materials: a centered 50~mm-diameter cylindrical tank of water (identified by the outer circle) and cylindrical inserts containing I-based, Gd-based, and Au-based contrast agents. The contrast inserts had concentrations between 0.2-1.6~mg/mL and were arranged in two rings. The outer ring contained single-contrast-agent solutions in water and the inner ring contained various two-contrast-agent mixtures. Figure \ref{fig:x_GT} shows the arrangement of contrast inserts as well as a Red-Yellow-Blue subtractive color-mixed image.

We simulated a diverging beam CT system with a source-detector-distance of 1105~mm, a source-axis-distance of 829~mm, 360 view angles, and 0.556~mm detector pixels. The spatial-spectral filter was positioned 380~mm from the source and was composed of Pb, Au, Lu, and Er tiles with thicknesses 0.25~mm, 0.10~mm, 0.13~mm, and 0.25~mm, respectively. Gantry rotation speed was 60 RPM and filter translation speed was 100~mm/s which corresponds to 0.5 pixels per view. We also modeled spectral blur due to filter motion and a 1.0~mm extended focal spot.

Poisson noise was added to the data scaled to $10^5$ (incident) photons per pixel per view for the lead-filtered spectrum. Reconstructions were performed using the model-based material decomposition algorithm described in Tilley et al. \cite{tilley2018model} with 0.5~mm voxels and using 800 iterations, ordered-subsets, and momentum-based acceleration. Material decomposition performance was characterized by the Root-Mean-Squared Error (RMSE) with respect to the ground truth for 4.0~mm cylindrical regions of interest (ROI) centered on each of the 6.0~mm cylindrical inserts. 

\FloatBarrier


\FloatBarrier



\vspace{-4mm}

\subsection{Filter Tile Order}

\vspace{-2mm}

To determine the effect of filter tile order, filter tile width was held constant at 1.52~mm (corresponding to an 8~pixel beamlet at the detector) and the order of the four filter materials was set to each possible permutation. Only the relative order of filter tiles was changed in this experiment, not the pattern phase shift (i.e. start position for filter translation). Therefore there were six possible filter tile orders for a set of four filter materials. Eight trials were also conducted for each filter order to characterize performance in the presence of noise. This experiment was designed to answer whether the joint effect of the spectral response of the filter materials and their relative position in the linear filter array would lead to variations in material decomposition performance.

\vspace{-2mm}

\subsection{Filter Tile Width}

\vspace{-2mm}

To determine the effect of filter tile width, filter order was held constant in descending order of K-edge energy (Pb-Au-Lu-Er) and the filter tile widths were adjusted between 0.2-15.3~mm corresponding to 1-80~pixel beamlets at the detector. Filter tile width changes the spatial-spectral sampling pattern as shown in Figure \ref{fig:spatialSpectralSampling}. 

If the filter tile width is too small, there is the potential for significant variation of the spectrum over the width of the beamlet due to blurring effects (e.g. extended focal spot). In  contrast, if the filter tile width is too wide, some regions of the image may be undersampled in some spectral channels. The motivation for this numerical experiment was to inform engineering design decisions by characterizing those regions of operation.

\vspace{-2mm}

\section{Results}

\vspace{-2mm}

\subsection{Filter Tile Order}

\vspace{-1mm}

\begin{wrapfigure}[6]{r}{0.64\textwidth}    
    \vspace{-16mm}
    \centering
    \includegraphics[clip,trim={2.0cm 0.0cm 4.0cm 1.8cm}, scale = 0.26]{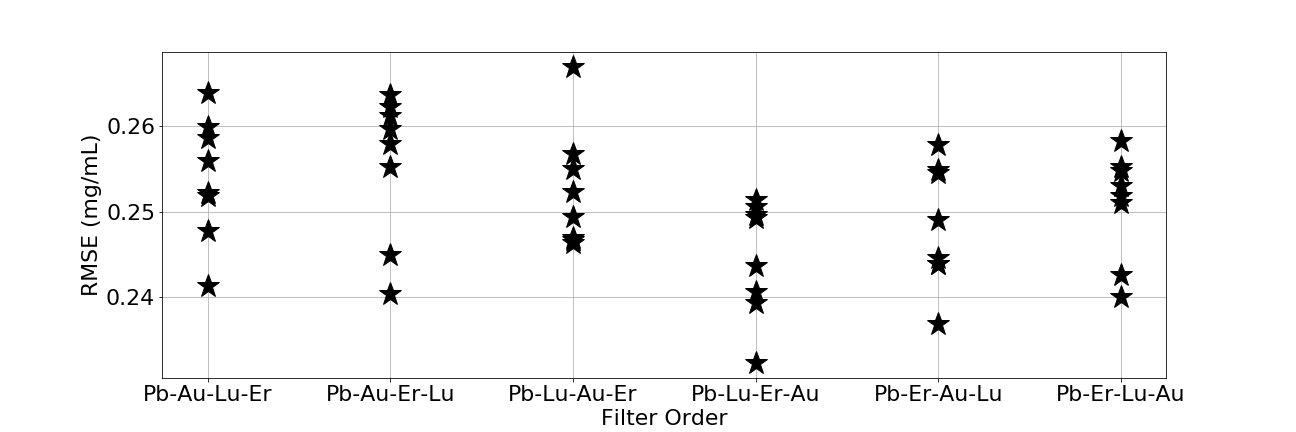}
    \caption{Filter order results for eight decompositions with noise.}
    \label{fig:filterOrderPlot} 
\end{wrapfigure}

Results for filter tile order experiment are shown in Figure \ref{fig:filterOrderPlot}. They show little, if any benefit to any of the six permutations. RMSE is centered between 0.24~mg/mL and 0.26~mg/mL for each case, a difference of less than 8\%, and the standard deviations of RMSE computed across the eight trials was between 0.008~mg/mL and 0.012~mg/mL, or around $\pm$5\%, for all cases. This result does not point to a specific filter tile order that is particularly beneficial or detrimental to performance.

\vspace{-3mm}

\subsection{Filter Tile Width}

\vspace{-2mm}

\begin{wrapfigure}[12]{r}{0.52\textwidth}
    \vspace{-10mm}
    \centering
    \includegraphics[clip,trim={0.8cm 0.4cm 1.8cm 2.2cm}, scale = 0.3]{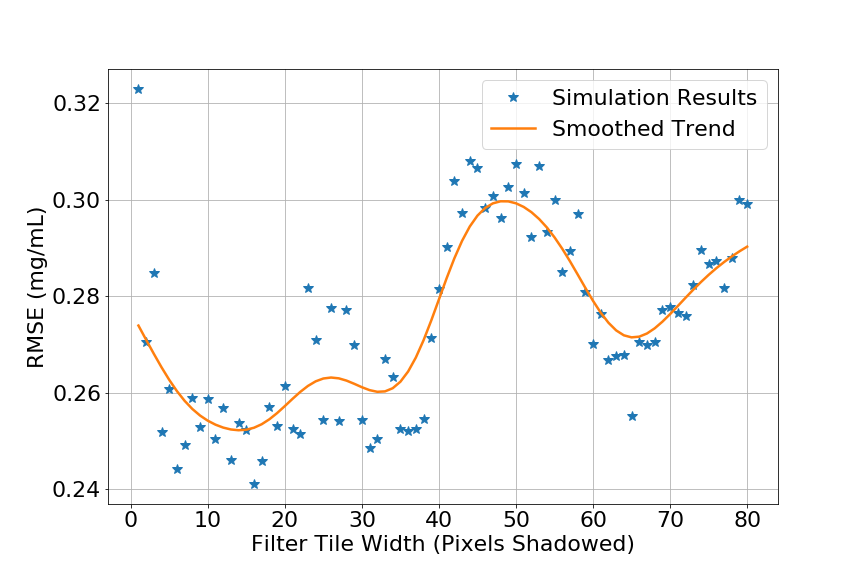}
    \caption{Filter tile width results. RMSE results in blue, gaussian-smoothed version in orange.}  
    \label{fig:filterTileWidthPlot}
\end{wrapfigure}

\begin{wrapfigure}[30]{r}{0.70\textwidth}    
    \centering
    \vspace{-14mm}
    \begin{subfigure}[t]{.35\textwidth}
      \centering
      \includegraphics[clip, trim={4cm 4.8cm 4cm 5cm}, scale = 0.15]{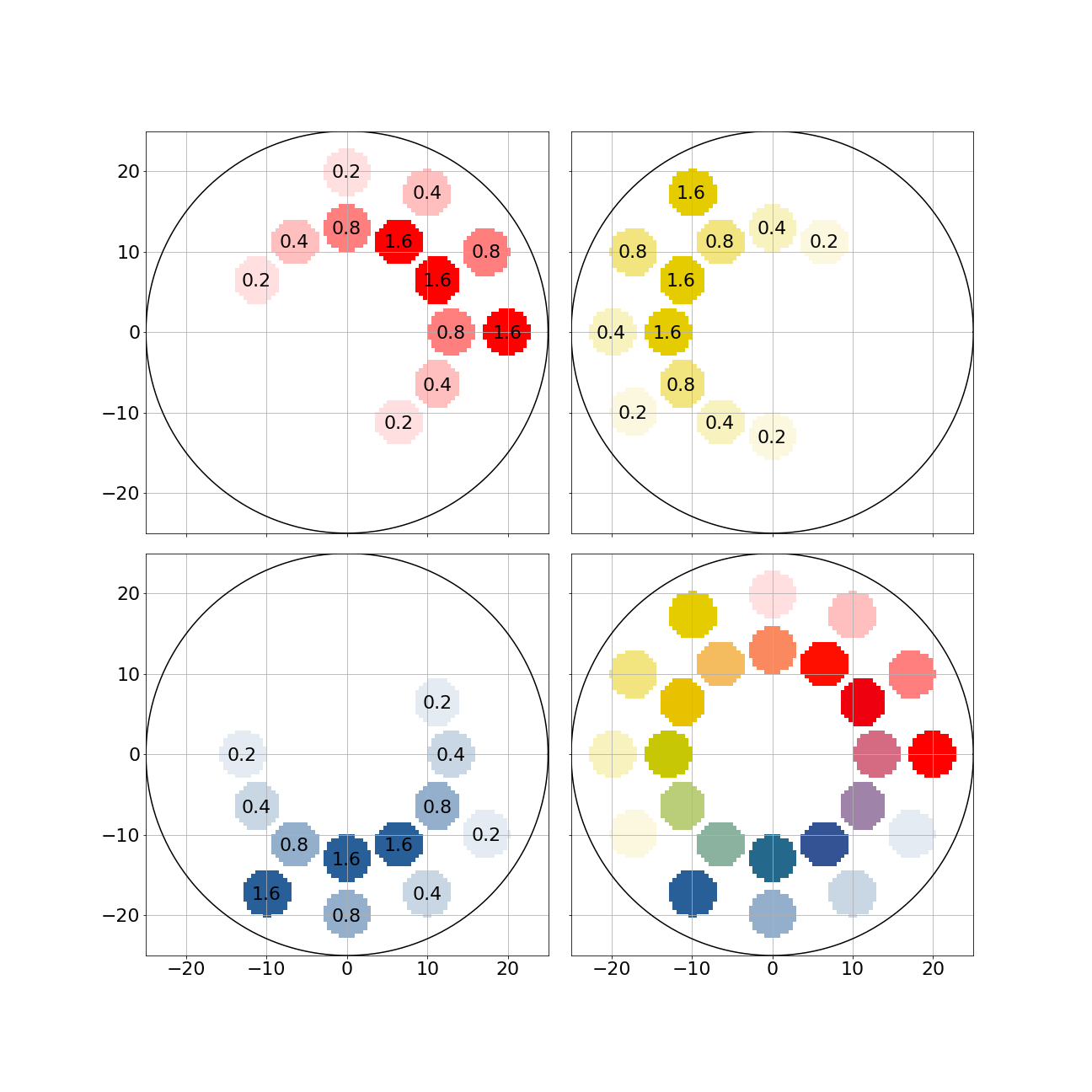}
      \caption{}
      \label{fig:x_GT}
    \end{subfigure}%
    \begin{subfigure}[t]{.35\textwidth}
      \centering
      \includegraphics[clip, trim={4cm 4.8cm 4cm 5cm}, scale = 0.15]{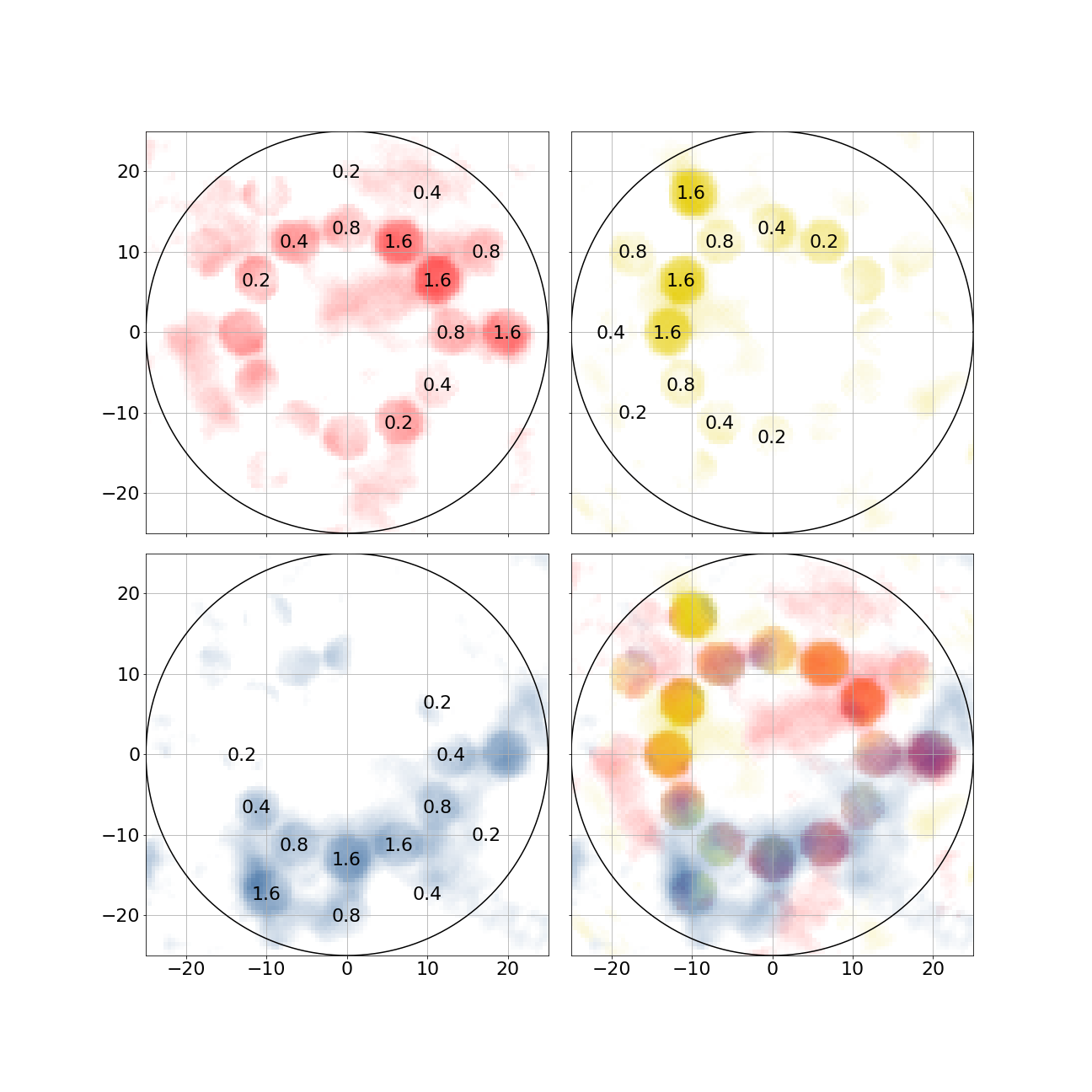}
      \caption{}
      \label{fig:xi_3}
    \end{subfigure}%
    \vspace{4mm}
    
    \begin{subfigure}[t]{.35\textwidth}
      \centering
      \includegraphics[clip, trim={4cm 4cm 4cm 5cm}, scale = 0.15]{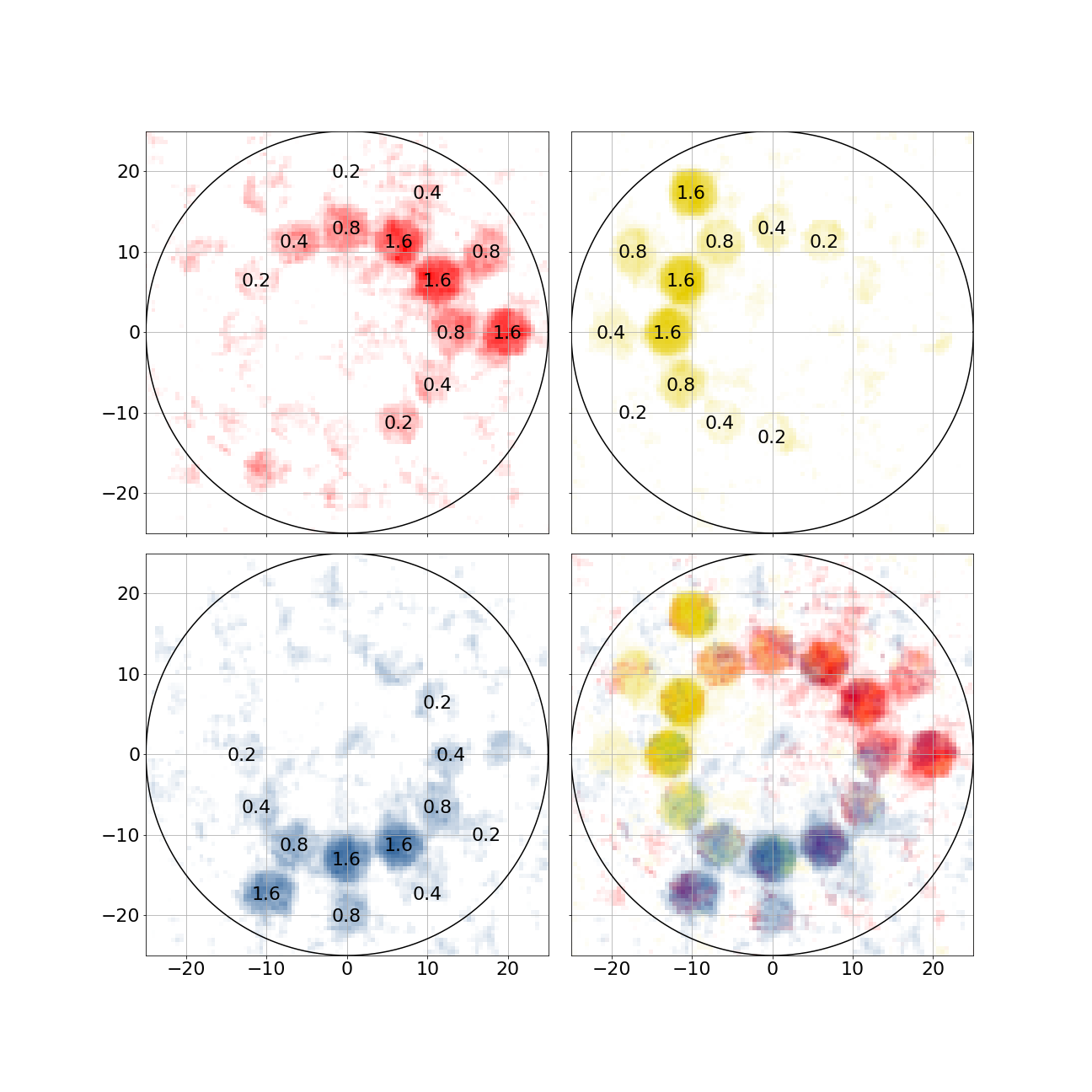}
      \caption{}
      \label{fig:xi_8}
    \end{subfigure}%
    \begin{subfigure}[t]{.35\textwidth}
      \centering
      \includegraphics[clip, trim={4cm 4cm 4cm 5cm}, scale = 0.15]{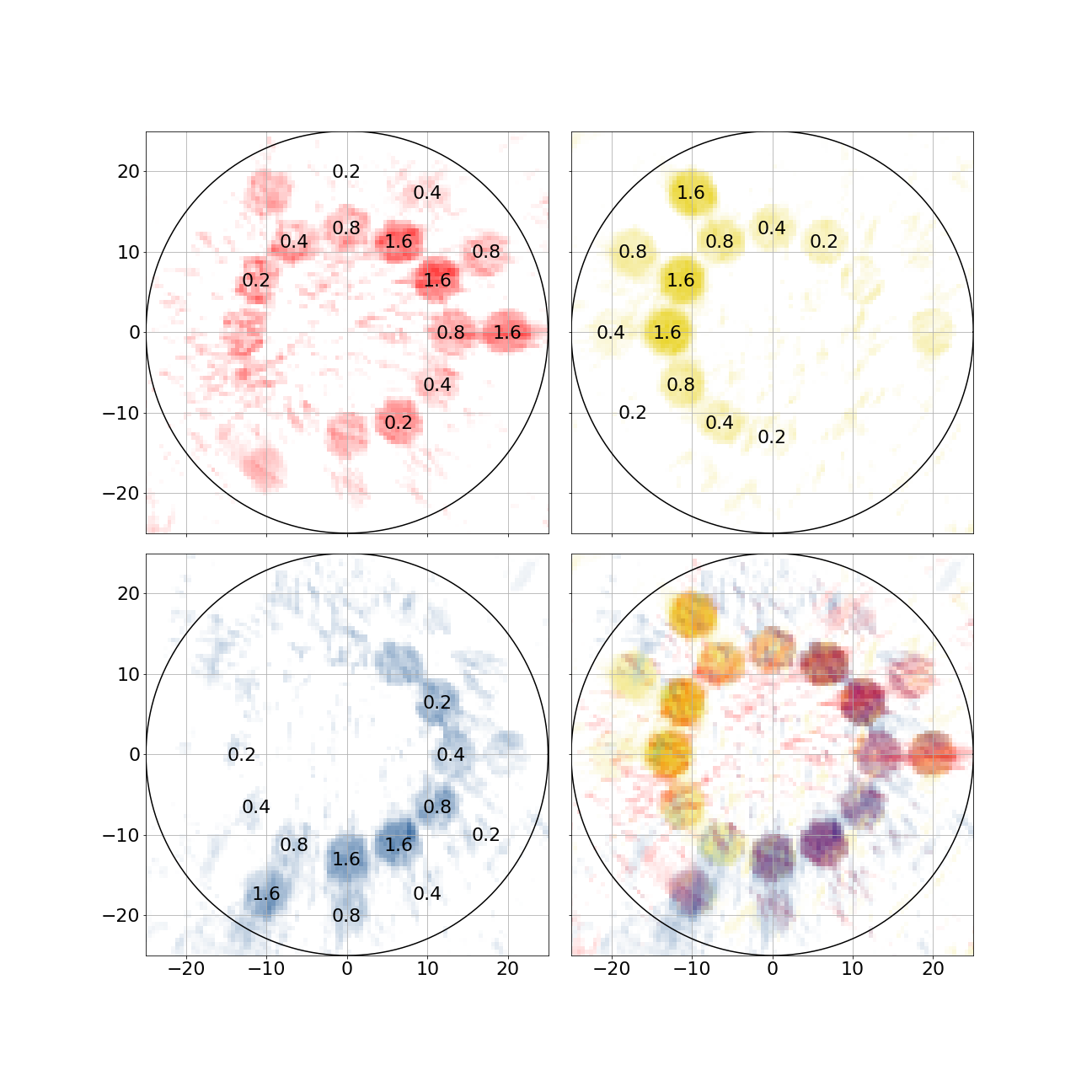}
      \caption{}
      \label{fig:xi_50}
    \end{subfigure}%
    \vspace{4mm}
    \includegraphics[clip, trim={0 0 0 0}, scale = 0.3, align=c]{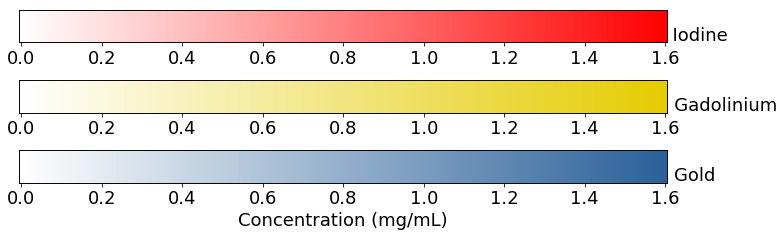}
     \includegraphics[clip, trim={4.5cm 3.3cm 2.5cm 3.3cm}, scale = 0.16, align=c]{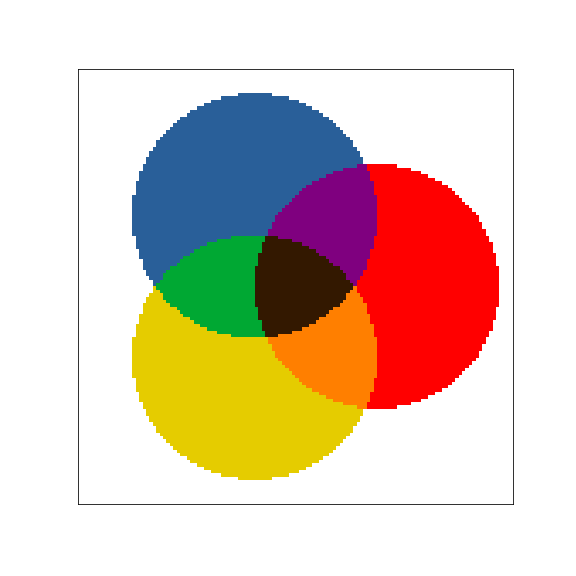}
    \caption{Reconstruction and material decomposition for (a) ground truth, (b) 3-pixel, (c) 8-pixel, and (d) 50-pixel beamlet width cases.}
    \label{fig:RYB}
\end{wrapfigure}

Image results for the filter tile width simulations are shown in Figure \ref{fig:RYB}. The images for the 3-pixel beamlet case in \ref{fig:xi_3} shows poor material decomposition performance, as does the 50-pixel case in \ref{fig:xi_50}. The closest match to the ground truth in \ref{fig:x_GT} is the 8-pixel case in \ref{fig:xi_8} which strikes the balance between the two extremes. Note specifically the iodine and gold (red and blue) regions which are only well-separated in the 8-pixel case. The 8-pixel case is fine enough for sufficient sampling without being subject to significant spectral blur.

The RMSE plot in Figure \ref{fig:filterTileWidthPlot} shows the same phenomenon. There is a sharp decrease in the range of 0-6~pixel beamlets (note that the focal spot is 5.2~pixels at the detector). There is a gradual increase in RMSE for wider filter tile widths that have coarser spatial-spectral sampling.

The trend in Figure \ref{fig:filterTileWidthPlot} also features an interesting modulation. We suspect this may be due to effects that phase in and out with different spatial-spectral sampling patterns (which is also a function of the filter translation speed). For example, there are some sampling patterns in which lines of response are more likely to be sampled by the same spectrum as their complementary ray. One would expect improved performance if these redundancies were minimized.

\vspace{4mm}



\FloatBarrier

\vspace{-4mm}

\section{Conclusion}


Filter tile order does not appear to have a significant impact on material decomposition performance indicating that it is a low-priority design consideration. Filter tile width has a more substantial effect.  Based on the results of this work, we propose that filter tile width for spatial spectral filters should be chosen as fine as possible while keeping the beamlet width wider than the size of the blur due to the extended focal spot and filter motion during integration. These constraints will be balanced with filter fabrication limitations for the final filter design. Results from these experiments will be used as part of a rigorous quantitative design and development of a new spectral CT system using spatial-spectral filters. Ongoing and future work will include investigations of spatial-spectral sampling patterns including the impact of different filter translation speeds and, possibly, non-linear filter motions.  

\vspace{5mm}

\acknowledgments

This work was supported, in part, by NIH grants R21EB026849 and F31EB023783.

\small
\bibliographystyle{spiejour} 
\bibliography{report}

\end{spacing}
\end{document}